\documentclass{sigchi}

\toappear{
Permission to make digital or hard copies of all or part of this work for personal or classroom use is granted without fee provided that copies are not made or distributed for profit or commercial advantage and that copies bear this notice and the full citation on the first page. Copyrights for components of this work owned by others than ACM must be honored. Abstracting with credit is permitted. To copy otherwise, or republish, to post on servers or to redistribute to lists, requires prior specific permission and/or a fee. Request permissions from Permissions@acm.org.\\
{\confname{IDHF 2014}}, November 25 - 26 2014, Kochi, Japan}

\usepackage{graphics} 
\usepackage{times}    
\usepackage{url}      
\usepackage{array,mathtools,multirow,setspace,subfigure,wasysym,url}
\usepackage{caption}
\usepackage{balance}

\DeclareGraphicsExtensions{.eps}

\makeatletter
\def\url@leostyle{%
  \@ifundefined{selectfont}{\def\UrlFont{\sf}}{\def\UrlFont{\small\bf\ttfamily}}}
\makeatother
\def\pprw{8.5in}
\def\pprh{11in}

\begin{document}

\title{Usability Evaluation of Dwell-free Eye Typing Techniques}

\numberofauthors{1}
\author{
  \alignauthor Sayan Sarcar\\
    \affaddr{School of Information Technology}\\
    \affaddr{Indian Institute of Technology Kharagpur}\\
    \email{mailtosayan@gmail.com}\\
}

\maketitle

\begin{abstract}
Dwelling is an essential task to be performed to select keys from an on-screen keyboard present in the eye typing interface. This selection task can be performed by fixing eye gaze on a key for a prolonged time. Spending sufficient amount of time on each key effectively decreases the overall eye typing rate. To address the problem, researchers proposed mechanisms, which diminish the dwell time. We conducted a within-subject usability evaluation of four dwell-free eye typing techniques. The results of \textit{first-time usability study}, \textit{longitudinal study} and \textit{subjective evaluation} conducted with $15$ participants confirm the superiority of controlled eye movement based advanced eye typing method (\textit{Adv-EyeK}) than the other three techniques.
\end{abstract}

\keywords{eye typing; usability study; dwell-free eye typing; text entry}

\category{H.5.2.}{Information Interfaces and Presentation: User Interfaces - Input devices and strategies}{}


\section{INTRODUCTION}\label{sec1}\vspace*{5pt}

In recent times, gaze-based interaction has evolved as an alternate user interaction modality. It has significant impact in the field of text entry in the last few decades~\cite{majaranta2002twenty}. The probable cause may be its similarity with traditional text entry methods, only by changing the controlling organ from finger to the eye. An advantage of the eye modality which may attract the developers is, with the same setup supporting able-bodied, we can develop systems for disabled users who are capable of visual interaction. These days, many applications support text entry through eye gaze~\cite{majaranta2004effects,miniotas2003symbol,urbina2007dwell}, also in the mobile environment~\cite{drewes2007eye}. 

Unlike other input modalities, eye gaze supports few commands (as eye is always moving and always activated, it is unnatural to hold the gaze for long time etc.) like eye movement, fixation, blinking and winking of an eye. In an eye-tracking setup, eyes can be attached with the mouse pointer with the help of an eye tracking device (i.e. camera) and gaze-tracking software. Eye gaze-based text composition can be performed in three ways namely a) direct gaze pointing or \textit{Eye typing}, b) eye gesturing and c) continuous writing~\cite{bee2008,majaranta2009text}. 

In \textit{Eye typing} method, users require to perform text entry through an on-screen keyboard. There, selecting a key from the keyboard can be performed by placing eye pointer for a slightly prolonged duration, called as \textit{dwell time}~\cite{majaranta2009text}. The other eye behavior used for selection is through \textit{eye blink} performed on the desired key button. The second method, i.e. \textit{Eye gesturing}, supports eye movements to draw a specific pattern (gesture) for selecting a character if that pattern matches. In \textit{Continuous eye writing} method, commands are activated based on the natural movement of gaze (like positioning the gaze into an overlay button area selects a character, etc.). It is supporting the users' natural gaze movements as the eyes are always activated and roamed constantly. Methods where switching off the gaze is not required (like continuous drawing with no lifting of pen) can be chosen as a suitable platform to implement the method (e.g. using \textit{Dasher} interface for gaze-based text entry, where always it is needed to select the characters from the character stream). \textit{Gaze-controlled continuous writing} method is different from {Eye typing} as it does not require the eye pointer to be continuously active. In \textit{Eye typing} method, rather after selecting a character, the eyes can stay which is not possible in other methods. 

In this research, we aim to develop \textit{Eye typing} methods as those can support dwell-freeness. Besides gaining popularity among the alternate text entry mechanisms~\cite{majaranta2009text}, eye typing creates a number of design issues which make it a unique technique with the own set of research problems.

One design principle of keyboard suitable for eye typing is decreasing the number of keys as well as the space between keys in the layout to save screen space~\cite{miniotas2003symbol}. On the other hand, larger size keys help the user to select them easily, even in a setup having less spatial resolution~\cite{hansen2001bringing}. In some cases, the large size of the interface is observed with the fewer number of keys accommodated in it, which inevitably takes larger screen space for a button. Thus, to optimize between eye movement, screen space and user comfort, an optimal size of the keys and space between keys needs to be decided. Before selecting the key, searching of it in the interface is an essential task which is significantly affected by visual stimuli presented in the interface such as color, orientation, shape, size, spatial frequency etc.~\cite{wolfe1994guided}. Another important objective of developing effective eye typing keyboard is to optimize the screen area utilization, especially for small display devices. {\v{S}}pakov and Majaranta~\cite{spakov_scrollable_keyboard} (see Fig.~1a), Panwar et al.~\cite{eyeboard2012} developed eye typing keyboards which save screen space. 

Minimizing speed-accuracy trade-off at different levels of cognitive complexity is one of the major concerns in dwell-based eye pointing~\cite{Zhang2010}. Large dwell time prevents users from false selections most of the time, as well as brings tiredness in their eyes~\cite{majaranta2007text}. On the other hand, shorter dwell time increases the chance of \textit{Midas Touch} problem (the classic eye tracking problem referring in our topic as wrong character selection from the keyboard through gaze pointing~\cite{jacob1991use}). As a result, it is difficult to conclude that shorter dwell time always produces better text entry rate with accuracy. The fixed dwell time also sets the maximum typing speed limit as the user has to wait for the stipulated time on each character button before selecting it. Majaranta and R{\"a}ih{\"a}~\cite{majaranta2007text} stated that most gaze typing evaluations were conducted with novices using a constant, fairly long dwell time ($450-1000$ ms). Wobbrock et al.~\cite{wobbrock2008longitudinal} used a short dwell time ($330$ ms) and achieved fair text entry rate ($7$ wpm). \v{S}pakov and Miniotas~\cite{vspakov2004line},  Majaranta and R{\"a}ih{\"a}~\cite{majaranta2007text} and Panwar et al.~\cite{eyeboard2012} studied dynamic adjustment of dwell time. Although the typing result of those systems were better, they reported \textit{delay} (participants committed that it was hard for them to change typing speed quickly as the system responded with a delay~\cite{vspakov2004}) and \textit{involuntary variation} (after selecting a key with less dwell time, users cannot move their eyes off the target. In turn, as the dwell time decreases, this adaptive adjustment becomes less convenient to the user~\cite{vspakov2004}) as critical problems. So, a trade-off still remains among dwell time, text entry rate and accuracy of the interface.
 
An effective way to increase the eye typing rate is to minimize dwell time. It has been observed by researchers that depending on the flexible cognitive complexity of users during the eye typing experiment, dwell time can be altered instead of fixing at a particular value~\cite{majaranta_dwelltime_2009}. Majaranta et al.~\cite{majaranta_dwelltime_2009} developed an interface which dynamically adjusts the dwell time. Their results revealed the effectiveness of the method in achieving faster eye typing rate. Further, researchers attempted to develop gaze-based interaction methods which can diminish dwelling task for selecting a character key. Effective execution of dwell-free methods can produce a moderate improvement of text typing rate. Urbina and Huckoff~\cite{urbina2007dwell}, Morimoto and Amir~\cite{Morimoto2010ContextSwitch}, Bee and Andr\v{e}~\cite{bee2008}, Kristensson and Vertanen~\cite{kristensson2012potential} (see Fig.~1b), Sarcar et al.~\cite{sayan2013} and Chakraborty et al.~\cite{tuhin2014} proposed dwell-free eye typing mechanisms.

\begin{figure}[!ht]
\centering
\subfigure[Scrollable keyboard (\v{S}pakov and Majaranta, 2008) ]{\includegraphics[scale=0.35]{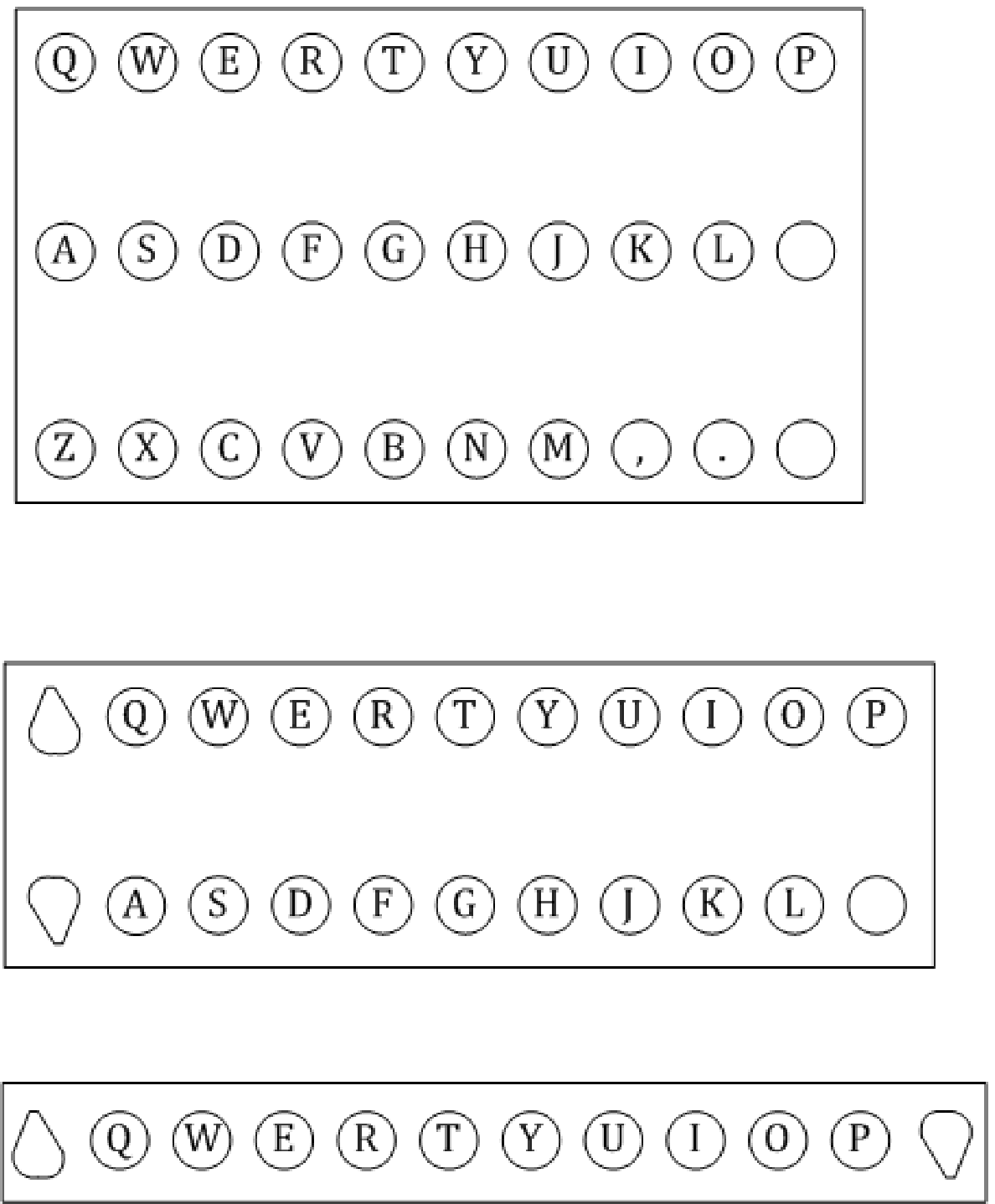}\label{fig:scrollable}} \hspace{1mm}
\subfigure[QWERTY keyboard (Kristensson and Vertanen, 2012)]{\includegraphics[scale=0.35]{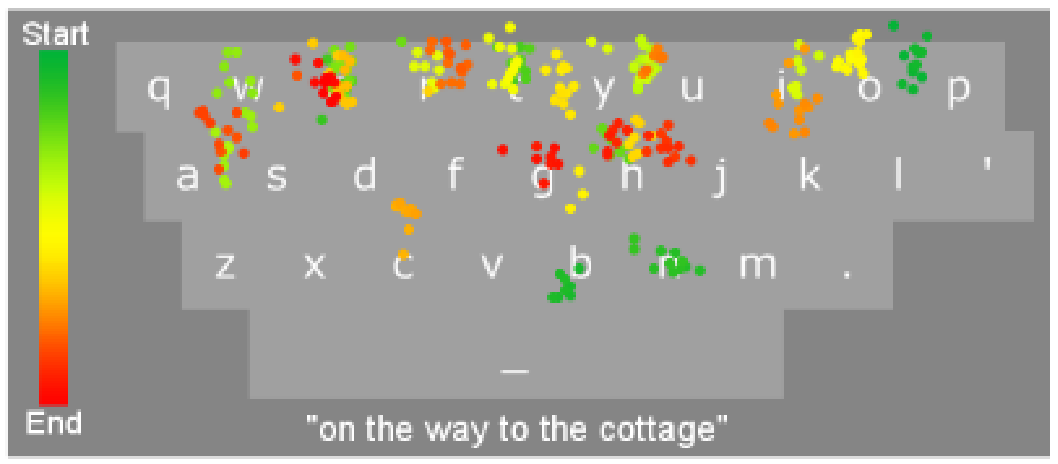}\label{fig:kristensson}}
\caption{Keyboards in eye typing interfaces}
\label{fig:diff_keyboards}
\end{figure}

According to eminent interaction designer Jacob Nielsen~\cite{nielsenusability}, \textit{usability} is defined as a quality attribute that assesses how easy and learnable a user interface is. The word ``usability'' also points to methods for improving ease-of-use during the design process. Our work seeks to examine the effectiveness of existing dwell-free eye typing techniques through two usability studies (a \textit{first-time usability study} followed by \textit{longitudinal study}) with respect to quantitative evaluation (i.e., eye typing rate and error rate) and subjective measures. The \textit{first-time usability study} is aimed to analyze the usability of popular eye typing methods/interfaces for short-time accessing. On the other hand, the objective behind performing \textit{longitudinal study} is to examine the combined effect of eye typing methods and on-screen keyboard layouts in accessing different eye typing interfaces of users over a long period of time. A subjective evaluation was performed using questionnaire responses collected during the experiment. From the results of usability between different dwell-free techniques examined in this comparative study, we find out the most effective technique which can further be used in developing efficient eye typing interfaces.

The organization of the paper is as follows. Section~2 surveys the state of the arts in dwell-free eye typing research area. First-time usability study and longitudinal study details and result analysis are provided in Section~3 and 4, respectively. Finally, Section~5 concludes the paper.

\section{Dwell-free Eye Typing Mechanisms}\label{sec2}
The detailed methodology descriptions of dwell time diminishing mechanisms considering for the usability evaluation are given below. 

\subsection{Iwrite}\label{subsec2.1}
\begin{figure}[!ht]
\centering
\includegraphics[scale=0.35]{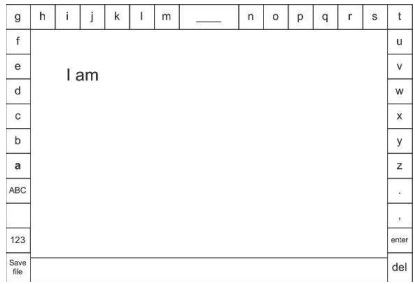}
\caption{Iwrite interface (Urbina and Huckauf, 2007)}
\label{fig:iwrite}
\vspace{6pt}
\end{figure}

Iwrite~\cite{urbina2007dwell} is a square-shaped interface for gaze-based text entry placing character buttons at the outer side and text area in the middle (Fig.~\ref{fig:iwrite}). The characters are selected by gazing toward the outer frame of the application. The text window is placed in the middle of the screen for comfortable and safe text review. The order of the characters, parallel to the display borders, reduces errors like the unintentional selection of items placed on the way to the screen button (e.g., \cite{ware1987}). The interface is very simple to use taking full advantage of the short saccade selection. 

\subsection{KKBoard}\label{subsec2.2}
\begin{figure}[!ht]
\centering
\includegraphics[scale=0.6]{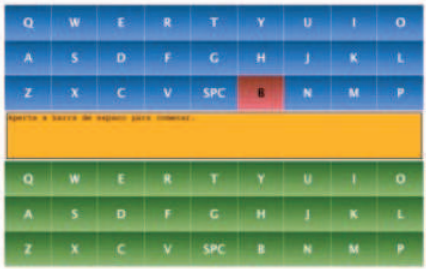}
\caption{KKBoard interface (Morimoto and Amir, 2010)}
\label{fig:kkboard}
\vspace{6pt}
\end{figure}

Morimoto and Amir~\cite{Morimoto2010ContextSwitch} proposed \textit{Context Switch} (CS) concept to activate a key selection. The CS concept is based on \textit{Key-focus} and \textit{Key-selection} task oriented eye movements. The \textit{KKBoard} interface replicates same keyboard layout placed in two separate screen regions, called as contexts (see Fig.~\ref{fig:kkboard}) where the user needs to switch alternately for selecting a key. In the eye typing process, key-focus task, activated through short dwell time, is followed by key selection, which is made by switching contexts (a saccade to the other context). The key which is last in focus in the previous context is taken at the selection phase. Context switching task diminishes the effect of the \textit{Midas Touch} problem. The CS-based text input replaces the traditional long dwell time with a short dwell followed by a context-switching saccade.

\subsection{EyeK}\label{subsec2.3}
\begin{figure}[!ht]
\begin{minipage}[c][6cm][t]{.25\textwidth}
  \vspace*{\fill}
  \centering
  \includegraphics[width=4.8cm]{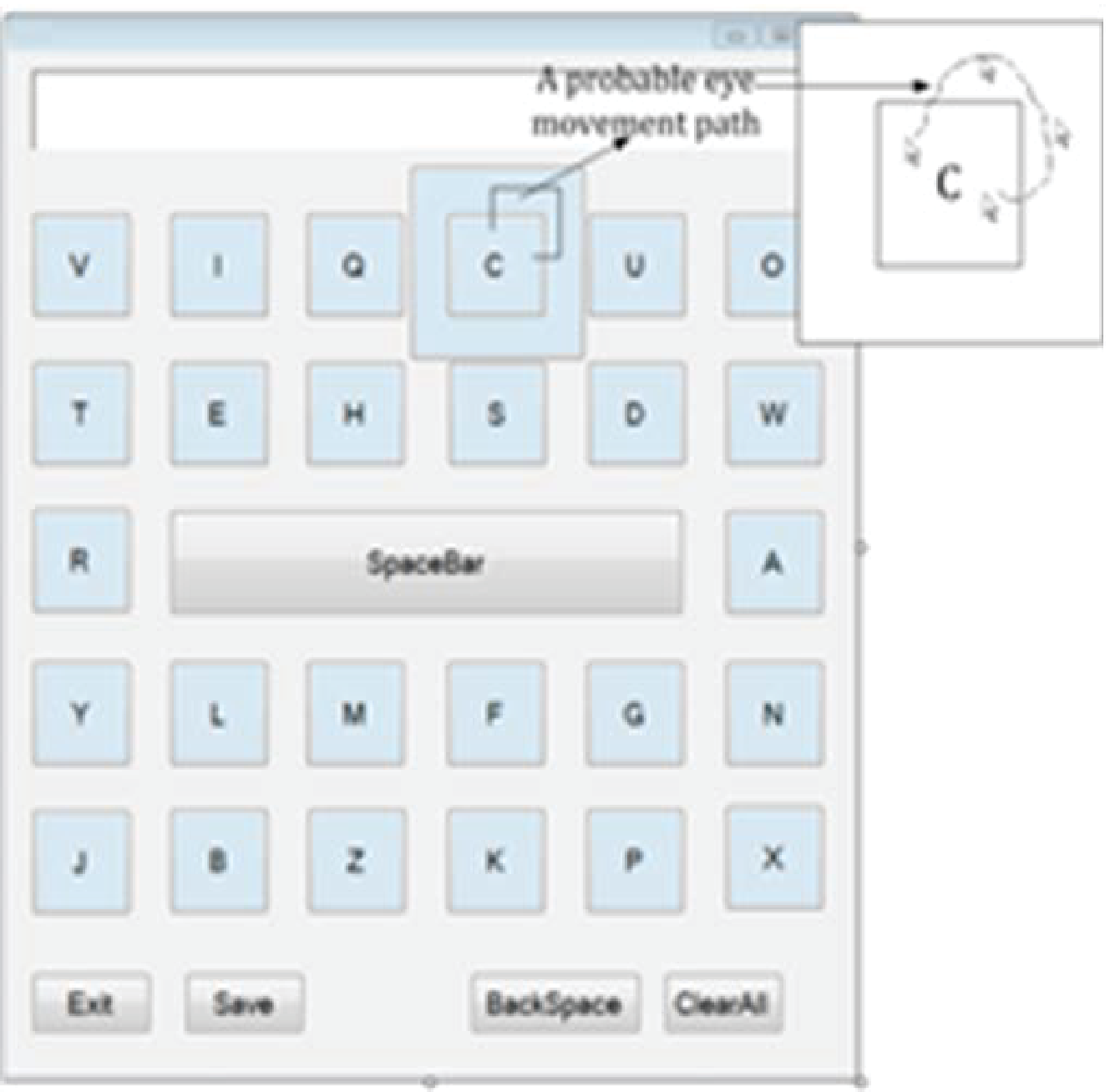}
  \captionof{subfigure}{Eye movement on a key `C'}
  \label{fig:eyek1}
\end{minipage}%
\begin{minipage}[c][6cm][t]{.25\textwidth}
  \vspace*{\fill}
  \centering
  \includegraphics[width=3.8cm]{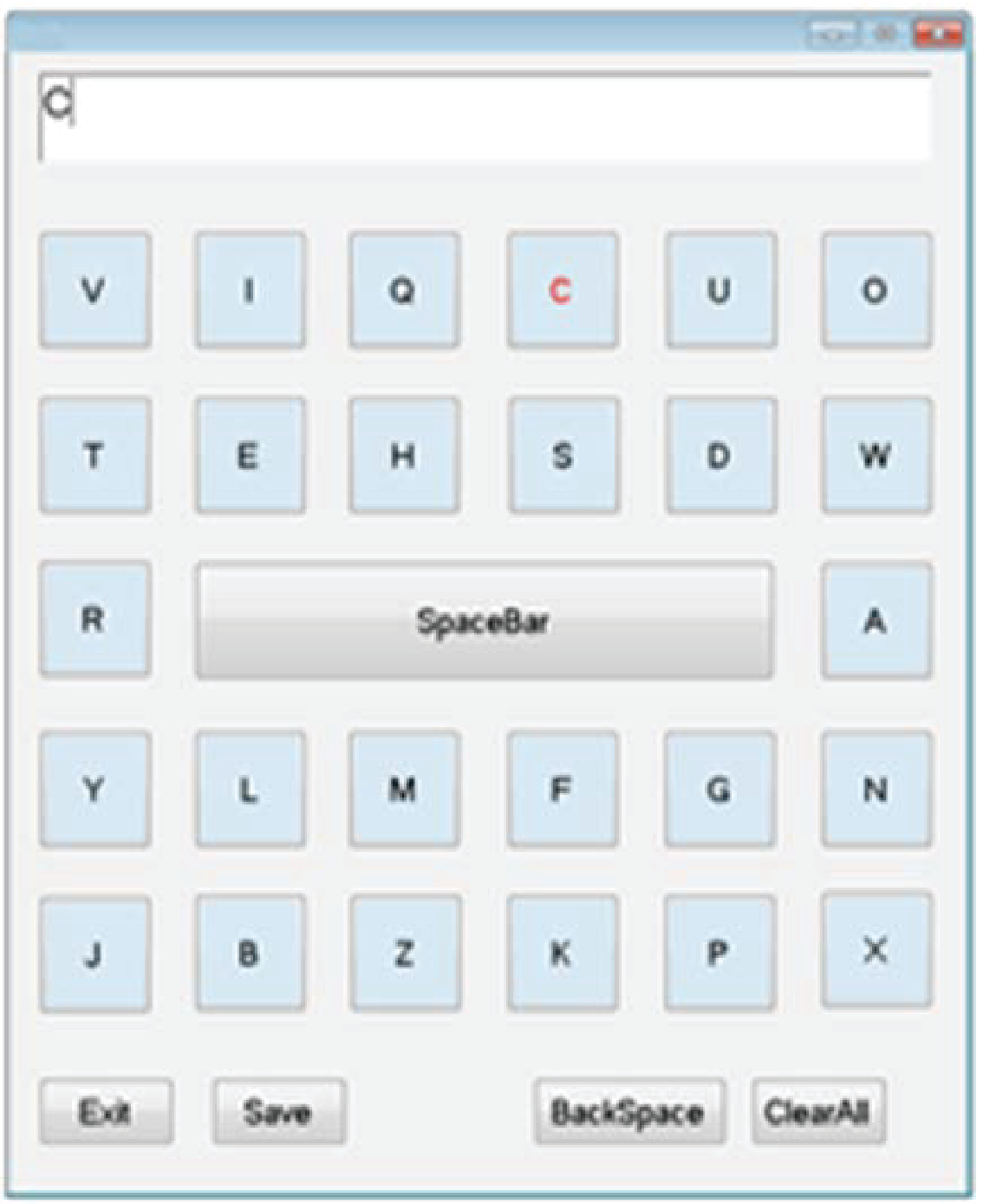}
  \captionof{subfigure}{`C' gets selected}
  \label{fig:eyek2}
\end{minipage}
\caption{Dwell-free eye typing in \textit{Adv-EyeK} layout (Sarcar et al., 2013)}
\label{fig:eyek}
\end{figure}

Sarcar et al.~\cite{sayan2013} proposed the \textit{EyeK} eye typing interface where they introduced a dwell-free typing mechanism. According to the method, a key button in the keyboard is selected automatically through specific interaction, which actually activates while a user moves the eye pointer through the button areas in inside-outside-inside fashion (see Fig.~\ref{fig:eyek}). As eye movement is faster than mouse or finger movement, this interaction takes minimum effort and time, which is negligible with respect to the traditional dwell time. According to the example shown in Fig.~\ref{fig:eyek}, suppose a user wants to select character `C' While hovering on the character, key and its overlay area are visible. The user starts moving the eye pointer within the key area, goes to outside overlay area and again comes back inside to the key area to complete the interaction phase (Fig.~4a, enlarged portion). After placing the eye pointer inside the key area, character selection gets activated (Fig.~4b). After selecting a character, visual feedback is given by changing its font color to red, which remains up to next character selection. If users need to enter same characters twice, they have to get out of the character initially and enter it again in the similar manner. This proposed mechanism provides a facility that the ``going out'' and ``coming back'' sides of a button may not be same and fixed. 

\subsection{An effective dwell-free eye typing technique (\textit{Adv-EyeK})}\label{subsec2.4}
\begin{figure}[!ht]
\begin{minipage}[c][6cm][t]{.25\textwidth}
  \vspace*{\fill}
  \centering
  \includegraphics[width=4.8cm]{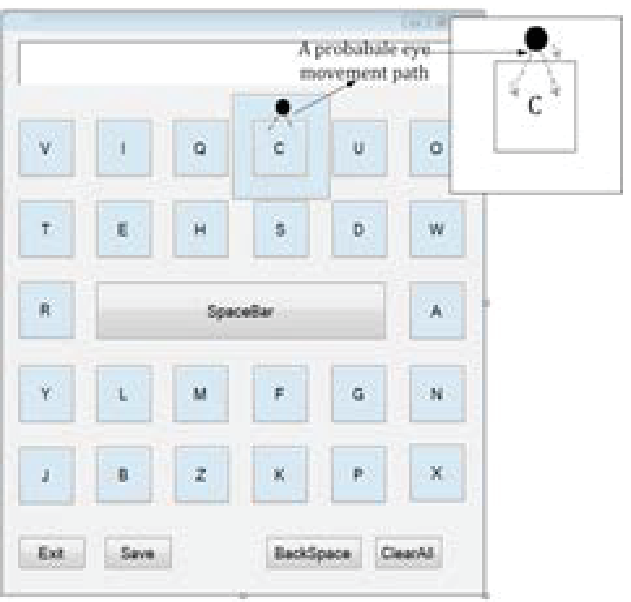}
  \captionof{subfigure}{Eye movement on a key `C'}
  \label{fig:chakraborty1}
\end{minipage}%
\begin{minipage}[c][6cm][t]{.25\textwidth}
  \vspace*{\fill}
  \centering
  \includegraphics[width=3.8cm]{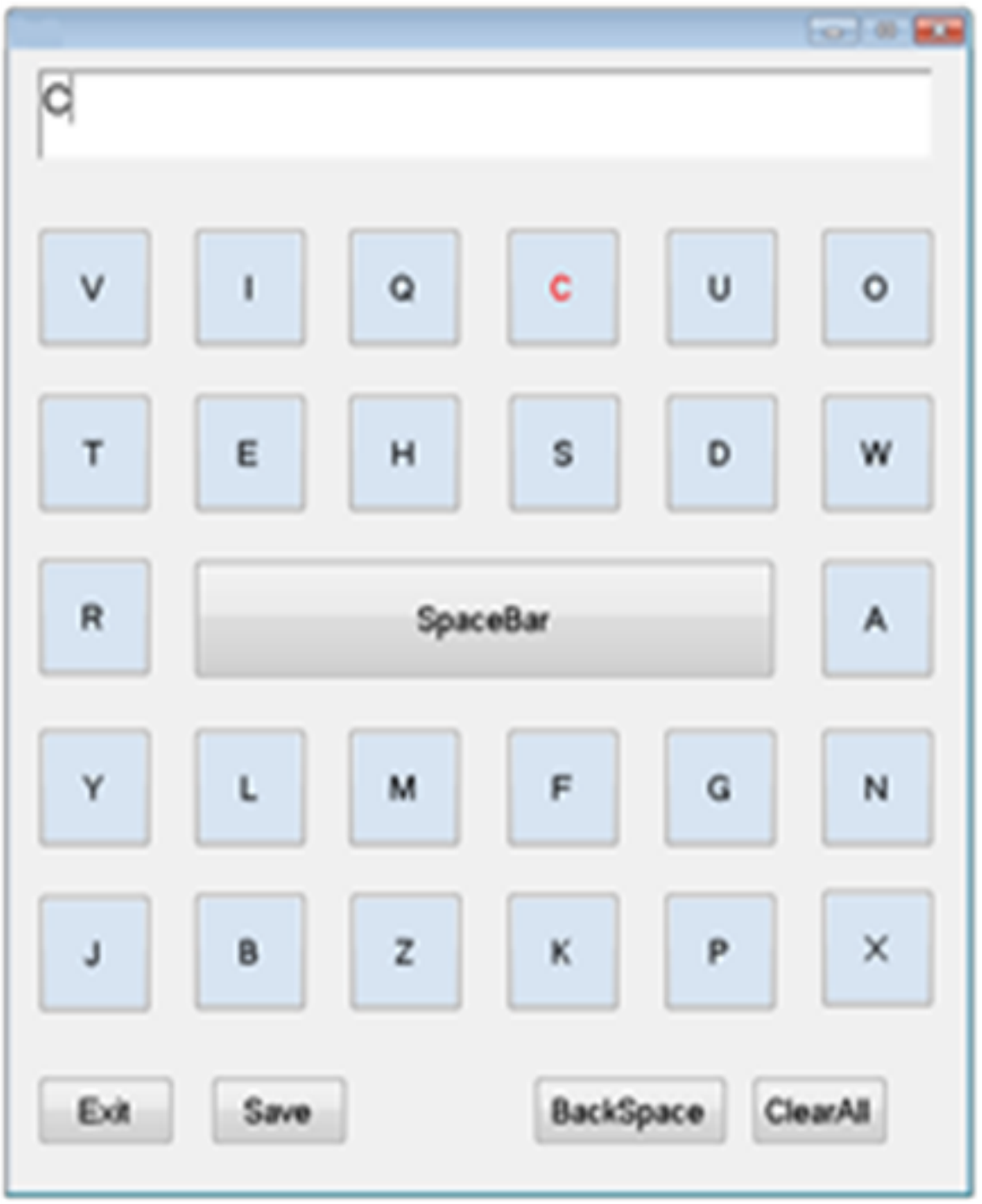}
  \captionof{subfigure}{`C' gets selected}
  \label{fig:eyek3}
\end{minipage}
\caption{Dwell-free eye typing in advanced \textit{EyeK} layout (Chakraborty et al., 2014)}
\label{fig:eyekk}
\end{figure}

A slight improvement over Sarcar et al.'s method~\cite{sayan2013} for dwell-free eye typing was proposed by Chakraborty et al.~\cite{tuhin2014} (throughout the article, we will call it as \textit{Adv-EyeK}, which is an advanced method of \textit{EyeK}). The interaction pattern for this advanced method is different from the previous method. It supports more controlled eye movement for key selection (Fig.~\ref{fig:eyekk}). In the overlay area which is activated after hovering, a black point is placed at a fixed position (upper side of the key, preferably at the center of the upper portion of the outer key area). After users hover on the intended key, they require to ``go out'' from the inner key, reach to that prominent point and after looking, ``come back'' inside the inner key area (preferably from same side) (Fig.~5a). A feedback system, same as the previous method, is applied to provide selection confirmation to the user (Fig.~5b). As an example, suppose, a user needs to select character 'C'. Then, on hovering `C', the outer layer becomes visible and user ``goes out'' from the upper side, sees the point and ``comes back'' into the inner key from the same side. For the double selection of a single key, the same procedure stated in the previous method is to be followed.

Kriestensson and Vertanen~\cite{kristensson2012potential} proposed a novel work in the related field, but they did not provide the working methodology of the system. So, we could not replicate the method. Bee and Andr\v{e}~\cite{bee2008} mimicked the design methodology of \textit{Quickwriting} text entry system for eye typing. We were unable to consider the system for our usability experiment also as it supports \textit{Eye Writing} instead of \textit{Eye typing}~\cite{bee2008}.  

We conducted two usability studies namely \textit{first-time usability test} and \textit{longitudinal user experiment} and analyzed the results. Our aim of conducting two esperiments was to fulfill two objectives: a) to observe the efficacy of popular dwell-free eye typing techniques with respect to user response in performing eye typing task over a short time period and b) to observe the user performance in a longitudinal study on combinations of on-screen keyboards suitable for eye typing with eye typing techniques.  
Detail descriptions of Design, Participants, Apparatus, Procedure, Results and discussions of both the usability studies are given below.

\section{First-Time Usability Test}\label{sec3}
The purpose of this usability test was to study the speed, errors and perception of usability in the first encounter with the eye typing interfaces. Conducting the short term study, we not only observed the initial reaction of users toward different eye typing interface, but also judged the efficacy of the interfaces with respect to immediate usability. We, in this study, wanted to observe the efficacy of different eye typing methods for short-term usability of users in different conditions (like accessing eye typing interface \textit{without} and \textit{with + without} help etc.). The other side of conducting this short study with every user is to provide scope to get acquainted with different eye typing interfaces which would further help them while performing long-term studies.   

\subsection{Apparatus}\label{subsec3.1}
\begin{figure}[!ht]
\centering
\includegraphics[scale=0.7]{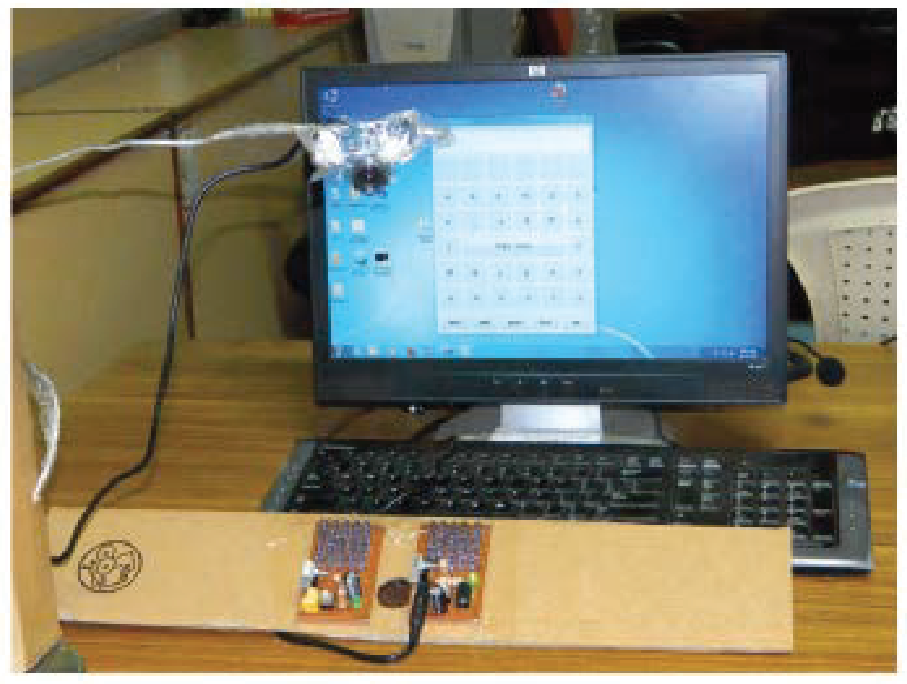}
\caption{Experimental setup}
\label{fig:expsetup}
\end{figure}

Experiments were conducted in a low-cost eye tracking setup using $2.2$GHz Intel Core2Duo processor with $15$'' screen LCD color monitor having $1440 \times 900$ resolution. Modified \textit{Sony PlayStation Eye} webcam, original lens was replaced by manual focus and Infrared (IR) filter removed lens, IR Lamp, consisting a matrix of $10$ IR LED, along with open source ITU GazeTracker software~\cite{SanAgustin2010ITUGazetracker}, developed by IT University of Copenhagen, were used for experiments (see Fig.~\ref{fig:expsetup}). The key press events and gaze positions were recorded automatically and stores into log files using separate event hooking programs. All experiments were performed in Windows $7$ environment. Controlled light conditions and positioning of the setup were maintained.

Three applications were run during testing the efficacy of $4$ dwell-free typing methods namely \textit{Iwrite}~\cite{urbina2007dwell}, \textit{KKBoard}~\cite{Morimoto2010ContextSwitch}, \textit{EyeK}~\cite{sayan2013} and \textit{Adv-EyeK} method~\cite{tuhin2014}. In each case, real-time eye movement was obtained with a tailored version of the application namely ITU GazeTracker. Apart from this, one C\# application was run to display eye typing interfaces. Another application was used for presenting phrases during the typing session running. This program randomly presented a target phrase from a corpus of $500$ to the user while simultaneously recording various text entry metrics for further analysis. 
Real-time (x, y) eye movements were converted to cursor coordinates by a C\# program that simply masqueraded the gaze point as the mouse position to other Windows applications. The program was based on ITU Gaze tracker developed at IT University of Copenhagen~\cite{SanAgustin2010ITUGazetracker}. 

Usability was measured by speed of text entry as words per minute (wpm) and total errors. In this study, we wanted to explore the effect of dwell-free eye typing mehods on users in short duration (in both with and with+without help situation). To judge user performance of traditional text entry systems, speed and accuracy were the basic and effective quantitative evaluation metrics. Thus, we collected user results on the basis of these two metrics.  

\subsection{Participants}\label{subsec3.2} 
Eye typing experiments were performed by $15$ participants ($11$ male, $4$ female) recruited from the local university area. Participants ranged from $25$ to $34$ years (mean = $28.5$). All were regular computer users, accessing on an average $4$ hours per day and have prior experience in composing text using eye typing techniques. All participants have normal or corrected-to-normal visual acuity. $14$ participants were right-eye dominant and $1$ was left-eye dominant, as determined from eye dominance test~\cite{eyedominancetest}. 

\subsection{Designs}\label{subsec3.3}
Three popular keyboards namely, a) compact screen space optimized full \textit{Scrollable Keyboard} layout proposed by \v{S}pakov et al.~\cite{spakov_scrollable_keyboard} (Fig.~1a), b) Panwar et al.'s key size and space optimized \textit{EyeBoard} layout~(Fig.~4b)~\cite{eyeboard2012}, incorporated in \textit{EyeK} interface~\cite{sayan2013} and c) popular QWERTY layout were considered for the evaluation. The size and space between key buttons were maintained same as specified in first two keyboards whereas for QWERTY, these were modified as $1.6$ cm for both height and width and $0.6$ cm for distance between two keys, respectively~\cite{sayan2013}. During the first-time usability experiments, sentences to be typed were taken from MacKenzie and Soukoreff's phrase set~\cite{mackenzie2003phraseset}. The experiment was a within-subjects design. Participants were randomly assigned to keyboard and eye typing method. 

\subsection{Procedure}\label{subsec3.4}
Every eye typing session was preceded by synchronization of the eye movement with the gaze tracker through \textit{Calibration}. During the typing session, users were instructed to type as fast as possible allowing few errors and not to move their eyes beyond the visibility range of the screen. To tackle this situation, users wrote the to be typed phrase using pen and paper before the session began or they listened to word by word from the instructor and composed them at runtime. Participants could only correct errors occurred due to wrong key selection by erasing them using backspace and then retyping them.

\begin{figure}[!ht]
\centering
\includegraphics[scale=0.6]{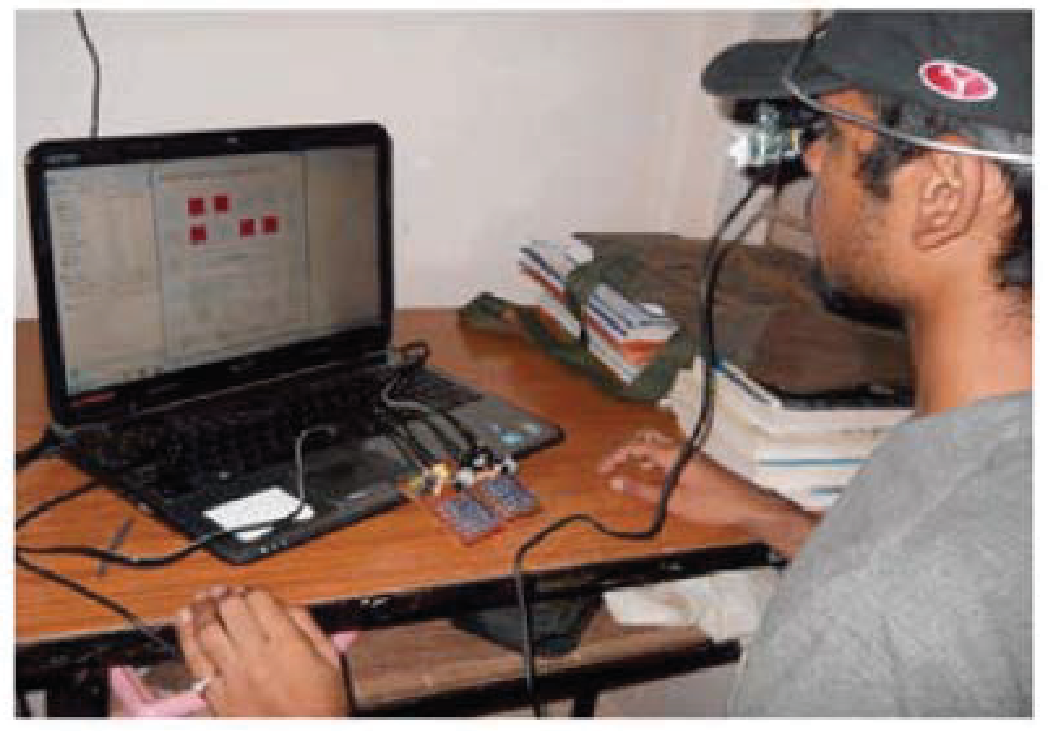}
\caption{Participants performing experiments}
\label{fig:partperexp}
\end{figure}

Users spent first few sessions before the experiment as training sessions where they were briefed about the nature of the experiment and then they completed a short demographic questionnaire. After that, users made themselves familiarized with eye tracking setup (fixing up camera and Infrared lamp positions) and eye typing methods applied on keyboard interfaces (Fig.~\ref{fig:partperexp}) by repeatedly accessing them. The total time for the interaction for each participant was about $10-15$ minutes. Before typing, participants first practiced some phrases on pen and paper, taken from standard phrase set~\cite{mackenzie2003phraseset} (this phrase set contains $500$ phrases, complete set is available in the web (link:{http://www.yorku.ca/mack/PhraseSets.zip}) to memorize them and then composed these phrases for each of the $3$ keyboards (to enhance the familiarity with keyboard based eye typing and typing texts). We collected feedbacks from the participants after the training session. If, after these practice sessions, participants felt the strong attachment with the keyboards and eye typing methods, then we started the testing sessions keeping a log for each.

In the testing phase of the first-time usability test, the initial speed of eye typing, errors, and perception of usability were studied. The first session of the initial usability test, on an average, took more than one hour, and data were not considered for analysis. Before starting of each session in short and long term test, users assured the instructors about their memorability of the practice set. On average, each testing session took about $45$ minutes. Among $12$ texts selected in each experimental procedure for a user, $2$ were taken from the in-domain Mackenzie and Soukreff's phrase set~\cite{mackenzie2003phraseset} and other $10$ were taken from out-of-domain texts such as novels, stories etc. for judging the design efficacy.

This test was performed to prepare the participants for longitudinal eye typing tasks with on-screen keyboards following specific typing methods within the constrained hardware scenario. At starting, an initial assessment of eye typing familiarity of the users was done. For this, in any of the keyboards chosen randomly, users were required to search some character keys given by instructor through eye movement and then select by fixing their eyes on the key for some time. This task was termed as \textit{first time test}. 
The eye typing experiment was planned such that participants were required to listen to the text and then typed it. Each user typed $20$ words in a session. In this context, we divided the session into two subgroups where the order became random for each participant. One was to type $5$ words, where we trained participants with single word typing after listening with no external help provided (\textit{without help} phase). Here, key selections were performed through dwelling. When the task began, instructor uttered one word, and the user was expected to type it with his eye as fast as possible. Once finished, he gave the signal to the instructor for supplying the next word. If a word was typed correctly, the system played a ``correct'' sound. If the word was typed wrong on the first attempt, the system gave a ``wrong'' beep rang once, and the instructor might either mark the error or repeat (predetermined) the help instructions. For the other method, known as \textit{with + without help}, each user was given to type $3$ phrase each containing $5$ words. Here, the instructor read the phrases once, user listened to that and typed as their own. Again, if users committed mistakes, instructors helped them to rectify those (this \textit{with + without help mode} was allowed only for two attempts per word or phrase. If the user got the second attempt wrong as well, the current word was skipped and he had taken to the next word). The aforementioned eye typing methods were conducted at least two times each for a user and ordering of the keyboard selection was counterbalanced across participants. Although taken from the same corpus, the testing words or phrases were different from those chosen in the training session. We allowed a maximum of two days gap between two consecutive sessions of a user.

\subsection{Dependent measures}\label{subsec3.5}
The dependent measures used in this experiment were words per minute (WPM) and the total error rate~\cite{soukoreff2003metrics,wobbrock2006erroranalysis} which is an addition of corrected and uncorrected error rate. 

\subsection{Results}\label{subsec3.6}

In the first-time usability test ($15$ participants $\times 20$ words), total attempts can be categorized into a) successes without help, b) the proportion of successes without help, c) successes with and without help, and d) the proportion of successes with and without help for each keyboard. We averaged individual category results with respect to an user. Out of $300$ attempts against each dwell-free methods, \textit{EyeK} and \textit{Adv-EyeK} method proposed by Chakraborty et al.~\cite{tuhin2014} earned best results in \textit{success with + without help} as well as in \textit{success without help} category. Using advanced \textit{EyeK} method, users achieved $215$ correct attempts (results ranging from $180$ to $230$, SD $= 1.15$) on an average which did not seek any help. The \textit{EyeK} interface also performed better as it achieved on an average $205$ attempts correct (range from $189$ to $223$, SD $= 1.2$) without seeking help out of $300$ attempts. The results achieved in this category by \textit{Iwrite} and \textit{KKBoard} interface augmented dwell-free mechanisms were $189$ attempts (range from $169$ to $203$, SD $= 1.21$) and $175$ (range from $156$ to $189$, SD $= 1.28$), respectively. It was observed that the impact of advanced \textit{EyeK} method was significantly better than other methods in terms of average success rate (F-test result, $F(3,292) = 5.92, p < 0.05$). In \textit{success \textit{with + without help}} category also, users preferred controlled eye movement which was supported by \textit{EyeK} augmented dwell-free method and advanced \textit{EyeK} based efficient method. Users got on an average $290$ correct attempts (results ranging from $285$ to $294$, SD $= 1.08$) and $296$ (range from $187$ to $205$, SD $= 1.05$) toward entering text through these methods. In contrast, the other two methods i.e. \textit{Iwrite} and \textit{KKBoard} interface associated dwell-free methodologies achieved only $239$ number of attempts (results ranging from $230$ to $249$, SD $= 1.07$) and $252$ (values from $238$ to $266$, SD $= 1.14$) \textit{with + without help} correct within $300$ attempts. In this category also, Statistical analysis revealed the significant difference of the newly proposed efficient dwell-free mechanism from other $3$ methods ($F(3,293) = 6.47, p < 0.05$). In this study, we observed no effect on different keyboard layouts while accessing dwell-free methodologies. 

During the user performing the usability task, we observed users' eye typing speed along with corrected (number of backspaces pressed) and uncorrected (error remaining in the typed text) error rate for all the methods. According to the scenario, all users typed in each keyboard augmented with every typing methodologies at least once. We gathered all user and keyboard based results of eye typing rate and error rate for a dwell-free eye typing methods and averaged them. The average eye typing rate of users in the first-time study ranged from $3$ to $6$ wpm. For \textit{without} as well as joint \textit{with and without help} based interaction, controlled eye movement based eye typing methods performed better. 

In case of \textit{without help}, \textit{Iwrite} and \textit{KKBoard} methods achieved average eye typing rate as $3.58$ (data varies from $3.2$ to $3.9$ wpm, SD $= 1.09$) and $4.14$ (data varies from $3.76$ to $4.36$ wpm, SD $= 1.19$) wpm, respectively. The \textit{EyeK} and \textit{Adv-EyeK} achieved text entry rate as $4.9$ (data varies from $4.2$ to $5.4$ wpm, SD $= 1.09$) and $5.6$ (data varies from $5.2$ to $6.0$ wpm, SD $= 1.08$) wpm, respectively. The statistical analysis on average eye typing rate reveals the existence of significant difference of the advanced method of \textit{EyeK} than other method's text entry performance ($F(3,293) = 5.94, p < 0.05$). The total error rate of the advanced method is 15.65\% (value ranged from $14.74\%$ to $16.25\%$ SD $=2.08$) which is lower than \textit{EyeK} (20.81\%) (value ranged from $18.74\%$ to $22.53\%$ SD $=3.12$) , \textit{Iwrite} (28.16\%) (value ranged from $26.78\%$ to $31.36\%$ SD $=4.12$) and \textit{KKBoard} (24.59\%) (value ranged from $22.14\%$ to $26.25\%$ SD $=3.48$). Also, there lies significant difference between total error rates of advanced method with other methods ($F(3,295) = 8.33, p < 0.05$). 
 
Analyzing the user results for \textit{with + without help} mode, we also found that the \textit{Adv-EyeK} method performed better in terms of text entry rate and total error rate than other eye typing methods. It achieved text entry rate of $5.5$ wpm (data varies from $4.8$ to $6.8$ wpm, SD $= 1.15$). The text entry rate of slightly different method associated with \textit{EyeK} interface got a text entry rate result as $5.35$ wpm (data varies from $5.2$ to $5.6$ wpm, SD $= 1.06$). The other two eye typing method associated with interface \textit{Iwrite} and \textit{KKBoard}, which are based on different principles than previous two, acquired text entry rate of $5.14$ wpm (data varies from $4.8$ to $5.4$ wpm, SD $= 1.09$) and $4.97$ wpm (data varies from $4.8$ to $5.1$ wpm, SD $= 1.04$), respectively. The statistical analysis on average eye typing rate in the \textit{with + without help} mode, reveals the existence of significant difference of \textit{Adv-EyeK} method than other method's text entry performance ($F(3,293) = 7.37, p < 0.05$). For the \textit{with + without help} mode based user evaluation, the total error rate of the advanced method is 11.65\% (value ranged from $10.37\%$ to $13.43\%$ SD $=3.0$) which is less than \textit{EyeK} (15.51\%) (value ranged from $13.04\%$ to $18.13\%$ SD $=3.8$) , \textit{Iwrite} (22.08\%) (value ranged from $18.78\%$ to $25.63\%$ SD $=4.00$) and \textit{KKBoard} (22.17\%) (value ranged from $20.05\%$ to $24.29\%$ SD $=3.00$). Also, there lies significant difference between total error rates of \textit{Adv-EyeK} method with other methods ($F(3,294) = 9.27, p < 0.05$).              

We observed total error rate (based on majorly uncorrected error rate because very few persons corrected the mistakes during composition with any keyboard), on an average over all the cases, of \textit{Adv-EyeK} method as 28.21\% whereas 33.52\%, 33.69\% and 31.89\% for methods augmented with \textit{Iwrite}, \textit{KKBoard} and \textit{EyeK} keyboard interfaces, respectively. We observed from the results that using the \textit{Adv-EyeK} method, users left fewer errors uncorrected. Further study on analysis of variance revealed that there was no significant difference in error rates between the methods (F(3, 295) = 1.03, n.s.).

\subsection{Discussion}\label{subsec3.7}
Results indicated that eye-typing with the \textit{Adv-EyeK} interface keyboard was significantly faster and yielded less error rate than three other eye typing methods namely \textit{Iwrite}, \textit{KKBoard} and \textit{EyeK}. The achieved high typing speed with the \textit{Adv-EyeK} method and low error rate can become more prominent with the number of sessions exercised by the users. Further, to explore the joint effect of eye typing method and augmented on-screen keyboard layouts, a longitudinal study was conducted.

\section{Longitudinal study}\label{sec4}

To analyze the effects on user performance on different dwell-free eye-typing schemes along with keyboards, we conducted a controlled experiment spanning 8 sessions attaching the methods with different eye typing keyboards including QWERTY. Using this many sessions allowed us to assess the performance of the $4$ methods over time, as we expected the learning rates of each method to be different. Individuals performed no more than $2$ sessions per day. maximum 48 hours gap was kept between two consecutive sessions for a user. If two sessions were performed on the same day, at least two hours were kept between sessions.

\subsection{Apparatus}\label{subsec4.1}
The apparatus matches that of the first-time usability test.

\subsection{Participants}\label{subsec4.2}
All the participants participated to first time usability tests were invited to participate in a longitudinal test. 

\subsection{Designs}\label{subsec4.3}
Same designs considered in the first-time test were used for this study. The study was a four-factor within-subjects design, with factors for Method
(\textit{Iwrite}, \textit{KKBoard}, \textit{EyeK} and \textit{Adv-EyeK}), Keyboard (QWERTY, \textit{Scrollable} and \textit{EyeBoard}) and Session (1-8).

\subsection{Procedure}\label{subsec4.4}
In the longitudinal test, it was decided that evaluating the performance of the participants could be judged by typing sufficient number of words through each of the three on-screen keyboard augmented with different eye typing methods, because it was observed that a typical user would not put much effort in small trials to learn typing. We divided the task into $8$ sessions for each user, each keyboard with each method.  While most sessions ran on consecutive days, we ensured that $2$ sessions could be completed for each user in a day. There could be a maximum gap of $2$ days between any two sessions for a user. No training was provided before the task. Users were asked to input approximately $8$ phrases ($40$ words) in a session as fast as possible. When the task began, the instructor spoke a phrase to the user which they listened and tried to type. No feedback, help or a second attempt was provided at the middle of typing a phrase. After completion of the phrase, participants pressed the ``save'' button, a sound was generated which alerted instructors for either declaring the session end or supplied the next phrases. Once the user had finished typing all the $8$ phrases, it was shown how well he has done (what he was supposed to type, what he typed on the first attempt, what he typed on the second attempt if any, the errors if any, and the speed). In this way, a user who wanted to complete all the sessions, he required to type at least $ 3$ keyboards $\times 4$ methods $\times 8$ sessions $\times 8$ phrases $= 768$ unique phrases taken from the standard phrase set. It was assured that instead of keyboard and method were selected randomly for a session, each user needed to type same $768$ phrases overall.  During the study, keyboard and method order were random and counterbalanced across participants. Each session lasted, on an average, approximately $45$ to $60$ minutes. A user, who evaluated all keyboards with different typing methods, approximately took $2$ to $4$ months to finish. Such a long duration is required to evaluate minimum $192$ ($2$ trials $\times$ $3$ keyboards $\times$ $4$ methods $\times$ $8$ sessions) trials per user. If the instructor, after a session, realized that user performance was not satisfactory, he conducted the session again. After a session completion, the user was asked to rate the task on a scale of 1 to 5 for difficulty on the basis of some demographic questionnaire. Then, the user was reminded to come for the next session before leaving.

\subsection{Dependent measure}\label{subsec4.5}
Aligned with the pilot study, the dependent measures used in this experiment were words per minute, overhead time, total error rate and subjective evaluaqtion parameters likeease-of-use, distracting, fatigue etc.

\subsection{Results}\label{subsec4.6}
Within-subject experiments were performed with $3$ keyboards each having $4$ dwell-free methods measuring \textit{eye typing rate}, \textit{total error rate} and \textit{overhead time}~\cite{kristensson2012potential}. Data for each participant were averaged in each session to form single measure per participant per session on a variety of metrics. Participants completed a total of $2$ trials $\times 3$ keyboards $\times 4$ methods $\times 8$ sessions $= 192$ trials. With $15$ participants, the entire study comprised of $2880$ trials. Also, for testing sessions, keyboard order was also kept counterbalanced across participants. $3$ sessions were performed per day by each participant. The whole study lasted for approximately $4$ months. Each trial was made of $8$ phrases taken randomly from new phrase set~\cite{vertanen2011} (this phrase set, developed by Vertanen and Kristensson~\cite{vertanen2011}, contains total of $2239$ sentences and sentence fragments generally used in E-mails; it can be accessed online (link:{http://aactext.org/comm2/comm2.zip}). 

\subsubsection{Eye typing rate}\label{subsec4.7}
\begin{figure}[!t]
\centering
\includegraphics[scale=0.45]{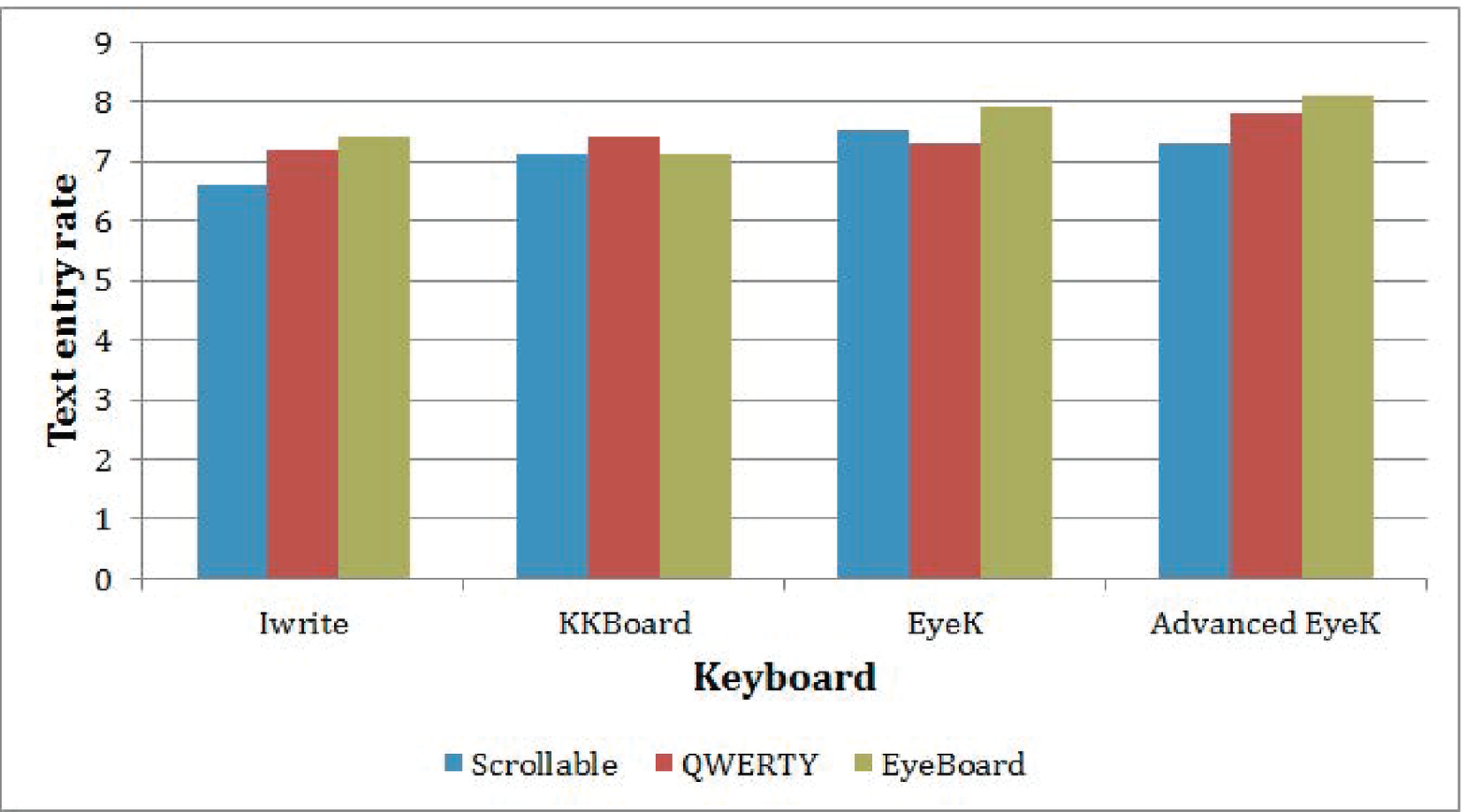}
\caption{Eye typing rate of different designs}
\label{fig:typingrate}
\end{figure}

The overall average eye typing rate achieved by participants with $4$ dwell-free methods applied on $3$ different keyboards was ranged from $6$ to $9$ wpm (Fig.~\ref{fig:typingrate}) (see Table~\ref{tab:tableeyetyping}). Using method applied in \textit{Iwrite}~\cite{urbina2007dwell}, keyboards \textit{Scrollable keyboard}, QWERTY and \textit{EyeBoard} earned the eye typing rate ranging from $5.9$ wpm to $7.1$ wpm (mean $= 6.6$, SD $= 1.05$), $6.2$ to $7.9$ wpm (mean $= 7.2$, SD $= 1.15$) and $6.4$ to $8.1$ wpm (mean $= 7.4$, SD $= 1.18$), respectively (Table~\ref{tab:tableeyetyping}). Implementing \textit{KKBoard} method, irrespective of users, $3$ keyboards achieved $6.4$ wpm to $7.9$ wpm (mean $= 7.1$, SD $= 1.02$), $6.5$ wpm to $8.2$ (mean $= 7.4$, SD $= 1.06$) and $6.55$ wpm to $7.8$ wpm (mean $= 7.1$, SD $= 1.17$), respectively. In case of \textit{EyeK} method, $3$ keyboards achieved $6.8$ wpm to $8.1$ wpm (mean $= 7.5$, SD $= 1.09$), $6.7$ wpm to $7.8$ (mean $= 7.3$, SD $= 1.06$) and $7.1$ wpm to $8.7$ wpm (mean $= 7.9$, SD $= 1.08$), respectively. Finally, participants on an average achieved eye typing rate of $6.7$ wpm to $7.8$ wpm (mean $= 7.3$, SD $= 1.07$), $7.4$ wpm to $8.3$ (mean $= 7.8$, SD $= 1.06$) and $7.3$ wpm to $8.9$ wpm (mean $= 8.1$, SD $= 1.08$), respectively through $3$ keyboards by using the \textit{Adv-EyeK} method. For all the mechanisms, it was observed that participants' eye typing rates got improved in the first few sessions and then reached to saturation. The analysis of variance (ANOVA) on text entry speeds showed that there was a significant difference between the means of user's performance on different eye typing mechanisms ($F(3,716$) = $5.42$, $p < 0.05$). Further, The Post-hoc using \textit{Tukey HSD} test reveals significant difference between performance of \textit{Adv-EyeK} and other methods ($p < 0.05$) (ANOVA or F-test can only tell about whether significant difference between the groups exist or not, but not which group significantly differs from the others. For that,  we need Post-hoc test; Tukey's HSD test is one of those tests which is commonly used). Also, we found no significant difference between the average performance of keyboard designs ($F(2,175) = 1.23$, n.s.).

\begin{table}[!t]
  \centering
  \caption{Average Eye Typing Rates}
    \begin{tabular}{|c|c|c|c|c|}
    \hline
    \multicolumn{1}{|c|}{\multirow{3}{*}{Methods}} & \multicolumn{3}{|c|}{Average Eye Typing Rate (wpm)} \\ 
    \cline{2-4}
    \multicolumn{1}{|c|}{} & \multicolumn{3}{|c|}{Keyboards} \\  \cline{2-4}
    \multicolumn{1}{|c|}{} & Scrollable & QWERTY & EyeBoard \\ \hline
    Iwrite & \multicolumn{1}{c}{6.6} & \multicolumn{1}{|c|}{7.2} & \multicolumn{1}{|c|}{7.4} \\ 
\hline
    KKBoard & \multicolumn{1}{c}{7.1} & \multicolumn{1}{|c|}{7.4} & \multicolumn{1}{|c|}{7.8} \\ 
\hline
    EyeK  & \multicolumn{1}{c}{7.5} & \multicolumn{1}{|c|}{7.3} & \multicolumn{1}{|c|}{7.9} \\
\hline
    Adv-EyeK & \multicolumn{1}{c}{7.3} & \multicolumn{1}{|c|}{7.8} & \multicolumn{1}{|c|}{8.1} \\
    \hline
    \end{tabular}%
  \label{tab:tableeyetyping}%
\end{table}%

\subsubsection{Overhead time}\label{subsec4.8}
Kristensson and Vertanen~\cite{kristensson2012potential} stated the task completion time, apart from the dwelling, as overhead time. In our case, as key selection time for large number of character entry took moderate time, key selection plus error correction time became overhead time. Throughout the sessions, we captured both key selection and error correction time and stored into a log file. The average overhead time for the $4$ methods, irrespective of keyboard and participant, were measured as 650 milliseconds (ms), 630 ms, 600 ms and 530 ms, respectively.

\subsubsection{Total error rate}\label{subsec4.9}
\begin{figure}[!t]
\centering
\includegraphics[scale=0.45]{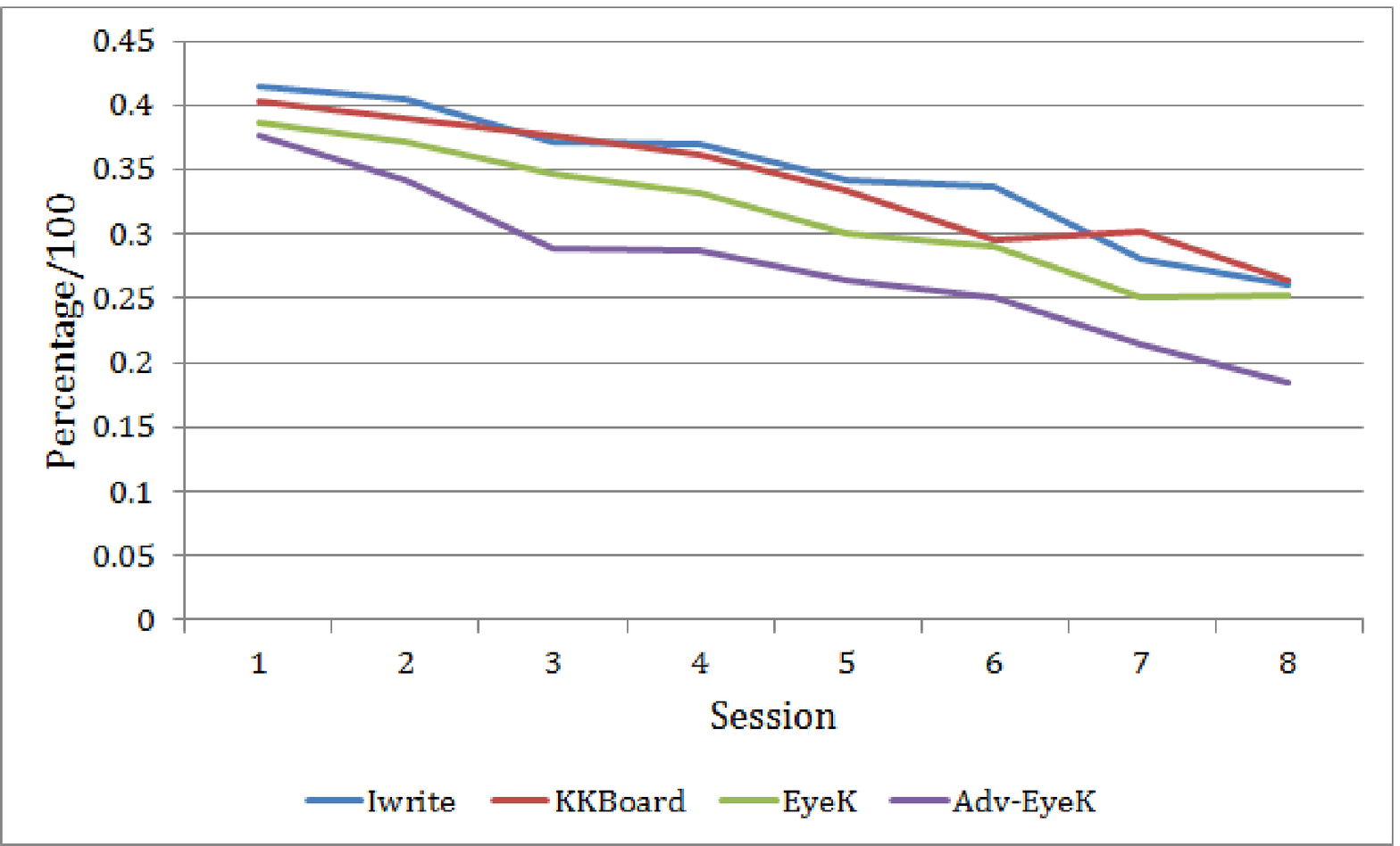}
\caption{Total error rate}
\label{fig:errorrate}
\end{figure}

Over the sessions, the total error rate, on an average over all the keyboards, became $26.29\%$ in case of \textit{Adv-EyeK} method and $33.15\%$, $33.18\%$ and $30.73\%$ for methods augmented with \textit{Iwrite}, \textit{KKBoard} and \textit{EyeK} keyboard interfaces, respectively (Fig.~\ref{fig:errorrate}). However, total error rates were dropped significantly over sessions ($F(7,350) = 4.29$, $p < 0.05$). The results concluded the observation that using the \textit{Adv-EyeK} method, users left fewer errors uncorrected in different keyboard designs, i.e., the number of corrected errors is more in case of associated interfaces. We also analyzed a number of errors left in the typed text for all the $4$ methods applied on $3$ keyboard designs. An analysis of variance revealed that there was no significant difference in error rates between the keyboard designs ($F(2,946) = 1.01$, n.s.)

\subsubsection{Subjective Evaluation}\label{subsec4.10}
We collected the subjective ratings from the participants with the \textit{nonparametric Wilcoxon Matched Pairs Signed Ranks Test}~\cite{wilcoxon2014} as Likert-scale data did not often conform to the assumptions required for ANOVA procedures. We talked with the participants before and after each session asking them about their eye strain and tiredness in a scale of $1$ to $5$. The level of tiredness was calculated by subtracting the first value from the later value. Analyzing the experimental results, we observed no significant difference between the average level of the tiredness, which was $0.52$ in the first and $0.71$ in the last session. We also calculated the text entry speed, ease of use, and general fatigue after each session using a questionnaire with a scale from $1$ to $5$. An increment of text entry rate of every user was observed (average from $3.1$ to $4.3$). By analyzing the participant preferences given in the Likert scale, \textit{Adv-EyeK} method performed significantly better than other eye typing techniques in terms of easier to use ($z = 42.00$, $p < .001$), faster ($z = 40.00$, $p < .01$), less distracting ($z = -47.00$, $p < .01$) and less fatiguing ($z = -55.00, p < .001$) (detail demographic questions and user feedbacks are shown in Table~\ref{tab:Servey}). 

\begin{table*}[!ht]
\renewcommand{\arraystretch}{1.4}
\centering
\scriptsize
\caption{Average scores for survey responses in the range 1-5}
\begin{tabular}{|l|c|c|c|c|}
\hline
Question   & Iwrite  & KKBoard  & EyeK   & Adv-EyeK                      \\ \hline 
How easy did you find the eye typing interface ?    &   3.31                   & 3.14                     & 4.08                      &  4.21        \\
(5- much easier, 1- much harder)                      &                      &                      &                                 &            \\ \hline
How tiring did you find using the dwell-free eye typing method ? &  3.12                    & 2.98                     & 3.91                      &  4.02          \\ 
(5- least tiring, 1- most tiring)                     &                      &                      &                       &                     \\ \hline
How faster did you compose the text ?                 & 3.22                     & 3.17                     & 4.07                      &  4.29  \\ 
(5- very fast, 1- very slow)                          &                      &                      &                       &      \\ \hline
How accurate was the composed text ?                  & 2.93                     & 3.02                     & 3.85                     &  3.96  \\
(5- accurate, 1- inaccurate)                          &                      &                      &                       &                      \\ \hline
How useful was the dwell-free eye typing system ?           & 3.32                     & 3.38                     & 4.14                      & 4.32 \\ 
(5- very useful, 1- less useful)                      &                      &                      &                       &                     \\ \hline
How distracting did you find the eye typing ?    & 2.55                     & 2.46                     & 2.13                      & 2.08  \\
(5- very distracting, 1- less distracting)            &                      &                      &                       &                     \\ \hline
\end{tabular}
\label{tab:Servey}
\end{table*}

Finally, participants were again interviewed after completion of the series of sessions. Participants felt that the concept of key area increment and allowing user's eye to move through those areas in a pattern to select an object (here, key button in an on-screen keyboard) was much more intuitive than other methods. They also admitted that typing by gaze was fairly easy, easier than their expectations and less boring than dwell-based eye typing methodologies, but clearly slower than using a conventional, hand operated hardware/virtual keyboard. 

Participants said that during first few sessions, the \textit{Adv-EyeK} and \textit{EyeK} interfaces were felt to be more difficult for them in terms of selecting an intended key than other dwell-free eye typing methods. Soon, they started to control their eyes easily and after some time, usually took hold of the methods and gradually geared up the speed with them (as well as felt comfortable). Although the \textit{Adv-EyeK} method was not known to participants previously, it can be pointed out as quickly learnable and perceived as easier to use than others. This might be for their simple going to a point and then coming back gesture of performing the selection task. Earlier, it might create a problem which soon got overcome. 

Regarding \textit{Iwrite} eye typing interface~\cite{urbina2007dwell}, participants expressed positive feedback about the simplicity and intuitiveness of it. But many of them said that they had difficulty to perform the interaction for selecting a key. 
The ambiguity between interaction and user's natural eye movement could hamper the eye typing process. 

Participants, after performing eye typing task with \textit{KKBoard} interface, admitted that in spite of the presence of popular QWERTY keyboard familiarity, the interface implemented two keyboard layouts which took double space than a normal keyboard size used in eye typing interface. Users suggested that the wastage of space should be minimized, but not that much where the keys become so small and inappropriate for use with eye trackers.
 
People might think that speed would also lead to greater perceived fatigue, but, in fact, from the observed results it was prominent that the \textit{Adv-EyeK} method was significantly less fatiguing than other methods executed through on-screen keyboards.

\subsubsection{Learning curve}\label{subsec4.11}
\begin{figure}[!ht]
\centering
\includegraphics[scale=0.32]{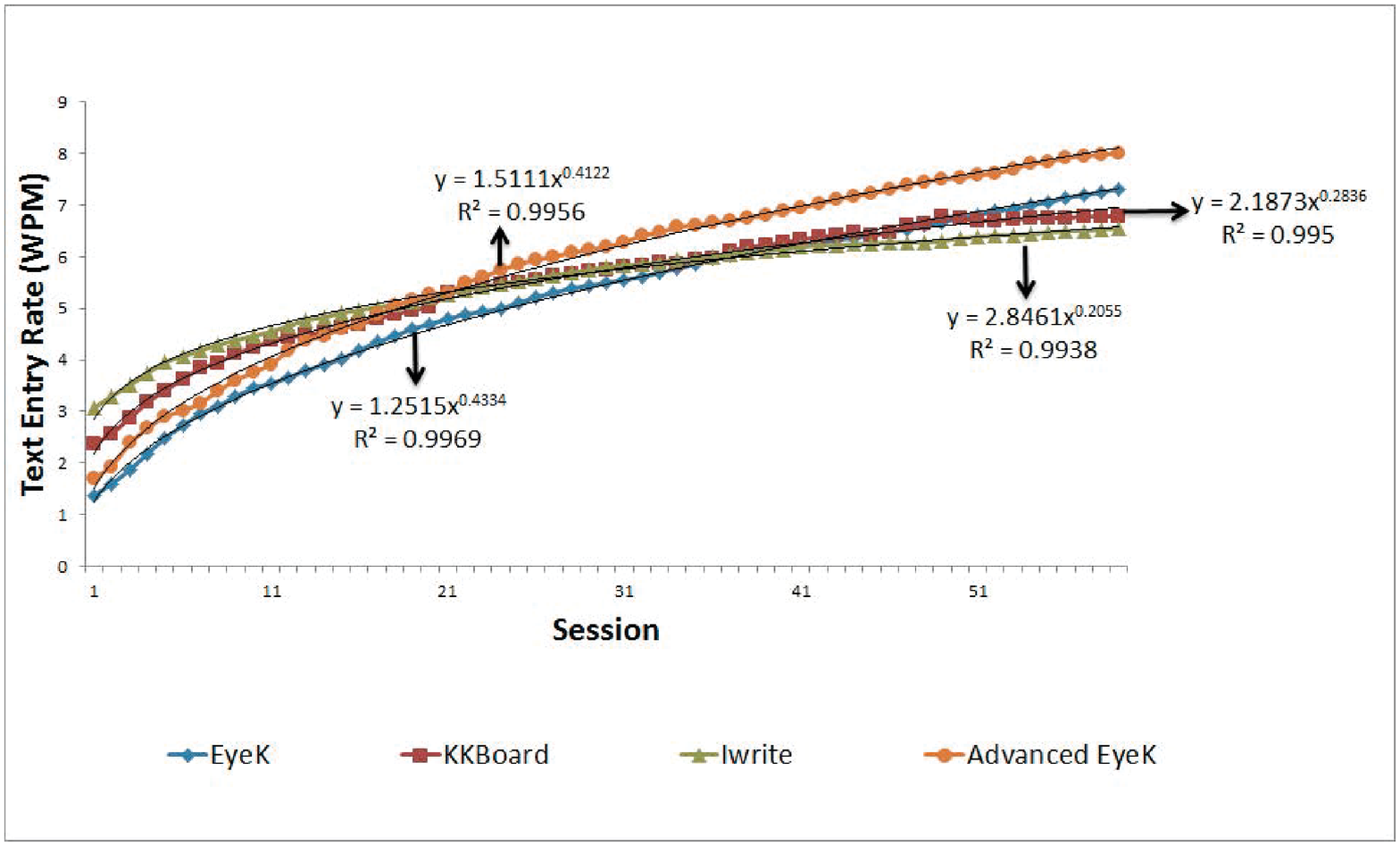}
\caption{Learning curve}
\label{fig:learningcurve}
\end{figure}

To understand the learning of the selected dwell-free eye typing methods augmented with on-screen keyboards we performed a longitudinal study with those $4$ methods augmenting QWERTY layout. $5$ new participants (who did not perform any of the previous experiments) having familiarity with QWERTY based text entry but unfamiliarity with eye tracking methods performed the eye typing sessions with testing phrases selected from Vertanen's phrase set~\cite{vertanen2011}. For each session, $5$ phrases had been typed by each user with each method. The average user result was depicted in Fig.~\ref{fig:learningcurve}. It indicates that the \textit{Adv-EyeK} method needed more initial effort to learn compared to other methods. However, after $20$ sessions, \textit{Adv-EyeK} outperformed other methods. We derived standard regression models in the form of the power curve fitting as it followed \textit{Power law of learning}. The longitudinal study lasted for $60$ sessions. The learning curve inevitably reflected the increasing efficiency of users after performing several sessions. The highest eye typing rate achieved through \textit{Iwrite}, \textit{KKBoard}, \textit{EyeK} and \textit{Advanced EeyK} methods were $6.55$, $5.79$, $7.31$ and $8.01$ wpm, respectively.

\subsubsection{Discussion}\label{subsec4.12}
Through user experiment, $4$ dwell-free eye typing methods were compared in terms of performance measures and subjective usability criteria. The first time usability and longitudinal study results reflected that \textit{Adv-EyeK} eye typing mechanism performed better in eye typing rate, error rate, and usability than other $3$ methods. The interface with on-screen keyboard implementing the method was learned quickly because of a fixed pattern based selection for every key. In view of users' subjective impressions, it is observed that they performed small eye pattern movement \cite{tuhin2014} well within a small region, after sufficient training. This scenario saves the screenspace offering an advantage over off-screen targets in limiting saccade distance to the dimensions of the attached keyboard's layout area. Further, subjective evaluations using standard task assessment tool \textit{NASA Task Load Index} (NASA-TLX)~\cite{hart1988} is underway and the preliminary observations assure the superiority of \textit{Adv-EyeK} design over others with respect to physical, mental and temporal workloads along with performance, effort and frustration measures.  

\section{Conclusion and Future Work}\label{sec5}
In this work, we present a usability study on $4$ dwell-free eye typing mechanisms comparing them with respect to eye typing rate, error rate, ease-of-use, eye fatigue etc. and collect results from short-term as well as longitudinal user experiments. The obtained user results reveal the suitability of \textit{Adv-EyeK} dwell-free eye typing method according to user's eye typing behavior. The learning rate and text entry rate of the \textit{Adv-EyeK} method augmented with on-screen keyboards are moderately higher than other eye typing techniques. This result undoubtedly support the superiority of the method even while attached with any of the popular on-screen keyboards applied for eye typing. Also, due to the controlled nature, the underlying interaction can accurately be performed by users, once become familiarized. 

In our study, we used our self-developed easy replicable low-cost eye tracking set-up, as one of the initial objective of this project was to develop low-cost eye typing task supporting hardware. Thus, we used low-cost apparatus and open-source software \textit{ITU gaze-tracker}.  However, the accuracy of the setup still is not up to the mark and thus, the validity of results confines within performing the experiments in controlled environments. To collect more accurate data from this low-cost setup, we introduced calibration before each session and also instructed participants not to see outside the screen area during the experiment. Irrespective of all these precautions, sometimes when we found that participants could not finish the experiment due to many errors committed, we had to conduct the same experiment again for each of them. Extending the current work, we can further improve the setup quality by many ways like fixing the infrared (IR) filters within visible range, placing the camera as close to eye for more accurately detecting eye gaze during calibration phase etc.

\balance

\bibliographystyle{acm-sigchi}
\bibliography{EyeBoard}



\end{document}